\documentstyle[twocolumn,aps,prl]{revtex}
\begin{document}
\hbadness=1500

\title{The quantum potential: the breakdown of classical symplectic symmetry and the
energy of localisation and dispersion}
\draft
\author{M. R. Brown}
\address{Department of Physics, Birkbeck College, University of London,
Malet Street, London WC1E 7HX, England}

\date{\today}
\maketitle
\begin{abstract}
The composition of the quantum potential and its role in the
breakdown of classical symplectic symmetry in quantum mechanics is
investigated. General expressions are derived for the quantum
potential in both configuration space and momentum space
representations.  By comparing the configuration space and momentum
space representations of the causal interpretation of quantum
mechanics, the quantum potential is shown to break the symplectic
symmetry that exists between these two representations in classical
mechanics.  In addition, it is shown that the quantum potential in
configuration space may be expressed as the sum of a momentum
dispersion energy and a spatial localisation energy; a
complementary expression for the quantum potential being found in
the momentum representation.   The composition and role of the
quantum potential in both representations is analysed for a
particle in a linear potential and for two eigenstates of the
quantum harmonic oscillator.

\end{abstract}


\section{Introduction}
\label{sec:intro}

Comparison of classical and quantum dynamics is not only of
interest at a fundamental level but is also of particular
conceptual and computational value in applications where both
classical and quantum phenomena play equally important roles (eg.
statistical mechanics, quantum chemistry, molecular dynamics etc).
However, such comparison can be hindered by the clear formal
differences between classical mechanics and the standard `operator'
formalism of quantum mechanics.

A common basis for comparing quantum mechanics and classical mechanics
was provided by the
introduction of the Wigner distribution function on phase
space~\cite{Wigner32}.  It brought to quantum mechanics not only
the advantages of C-number only
equations but also the appeal of classical concepts that accompany
a phase space description.
Thus, for example,
has the framework of quantum hydrodynamics (QHD) emerged from
zeroth, first and second momentum moments of the Wigner distribution
and been applied to semi-conductor device modelling~\cite{Ferry93,Gardner96}.
In such approaches, separable quantum internal energy
terms appear alongside classical
thermal pressure and scattering terms.  The explicit appearance
of an internal energy (or stress)
is a result of the projection from the phase
space representation to the configuration space representation.  This
point was made for pure states in quantum mechanics by
Takabayasi~\cite{Tkbysi53} and, more recently, has been
demonstrated by Muga et al.~\cite{Muga93} for classical mechanics and for
pure and mixed states in quantum mechanics.

As noted in~\cite{Muga93}
and exemplified
in~\cite{Ferry93,Gardner96}, the quantum contribution can be different in
different formalisms and/or approximations.  Ferry and
Zhou~\cite{Ferry93} draw attention to the lack of consensus as to the
form of the correction potential to be included in the QHD
equations used for semi-conductor device modelling.  They point out that both
the Bohm potential and the Wigner potential have appeared both
separately and together in various formulations of QHD.  (See references
in~\cite{Ferry93}).  Indeed, in~\cite{Ferry93}, Ferry and Zhou find a
correction term (hereafter referred to as the Ferry-Zhou potential)
which is the difference between the Bohm potential and the
Wigner potential as defined in~\cite{Ferry93}.
Implicit in this finding is the observation that the
Bohm (quantum) potential is expressible as the sum of two other
potentials, at least for a pure quantum state.  We make the further
observation that little if any attention has been given to the
appearance of a
quantum potential in the momentum space representation.

The last two observations motivate the present investigation of the
composition and role of the (Bohm) quantum potential in quantum
mechanics at a more fundamental level. However, rather than
starting from a phase space formulation of quantum mechanics, we
start from the the causal interpretation of quantum mechanics on
configuration space. This interpretation offers an alternative
basis for direct comparison of quantum and classical dynamics
through the Hamilton-Jacobi formalism. In the causal
interpretation, the quantum potential, in addition to the external
potential, guides the trajectory of the `quantum particle'
~\cite{book:bh93}.  The functional form and the effects of the
quantum potential in this interpretation have been extensively
studied for a wide variety of specific cases
~\cite{dh82,dkv84,dewdney85,dhk86,dhk87,dewdney87,ld90} and in
~\cite{book:bh93}.

In this paper, we give equal attention to the expression of the
causal interpretation and the appearance of the quantum potential
not only in the configuration space representation but also in the
momentum space representation. Rather than using a phase space
projection to obtain the momentum representation, we use the
momentum space representation of the position and momentum
operators to obtain the momentum space representation of the
quantum Hamilton-Jacobi equations.  A comparison of the causal
interpretation in its respective configuration space and momentum
space representations provides insight into the classical
symplectic symmetry breaking role of the quantum potential.

If the quantum potential reflects the quantum aspects of a system,
it should be possible to identify such aspects within the quantum
potential.  The quantum stress (an energy per unit volume in
configuration space) identified by Takabayasi~\cite{Tkbysi53}, is a
measure of momentum dispersion in configuration space and is
proportional to the Wigner potential mentioned above.  In this
paper, we show how in configuration space the balance of the
momentum dispersion energy in the Bohm quantum potential is formed
by the quantum localisation energy which is the Ferry-Zhou
potential also identified above.  The latter is proportional to the
curvature of the probability density; hence its description as a
localisation energy. We also establish the corresponding
relationship between the spatial dispersion energy and momentum
localisation energy in the momentum space representation.  Thus, we
find that the balance between localisation and dispersion energies
suggests a link between the quantum potential and the Heisenberg
uncertainty principle.

The paper is organised as follows.  In section~{\ref{sec:qpcuas}},
the appearance of the quantum potential in the standard
configuration space representation of the causal interpretation of
quantum mechanics is briefly reviewed. In
section~{\ref{sec:gender}}, the derivation of the quantum potential
is generalised in configuration and momentum space and its
emergence from the actions of the respective kinetic and potential
energy operators is demonstrated. In section~{\ref{sec:symplc}},
the symplectic structure of the causal interpretation is
investigated by expressing its equations in Hamiltonian form.  The
classical symplectic symmetry breaking property of the quantum
potential is demonstrated. In section~{\ref{sec:disploc}}, the
quantum potential is expressed as the sum of dispersion and
localisation energies and a connection with the Heisenberg
uncertainty principle is suggested. In section~{\ref{sec:expls}},
we examine the role and composition of the quantum potential for a
particle in a linear potential and for the ground state and second
excited state of the harmonic oscillator. We conclude in
section~{\ref{sec:conc}}.

\section{Quantum Potential in the Causal Interpretation}
\label{sec:qpcuas}

The quantum potential emerges from Bohm's Causal
Interpretation~\cite{bohm52} of quantum mechanics on configuration space,
when the substitution $\psi= Re^{iS/\hbar}$ is made in the
Schr\" {o}dinger equation:
\begin{equation}
i\hbar\frac{\partial\psi}{\partial t} =
\left(-\frac{\hbar^{2}}{2m}\nabla^2 + V \right)\psi \label{eq:schro}
\end{equation}
and the real and imaginary parts are separated yielding the equations:
\begin{equation}
\frac{\partial S}{\partial t} +
\frac{(\nabla S)^{2}}{2m} -
\frac{\hbar^{2}}{2m}\frac{\nabla^{2}R}{R} +
V = 0 \label{eq:hj}
\end{equation}
and
\begin{equation}
\frac{\partial\rho}{\partial t} +\nabla\cdot(\rho{\frac{{p}}{m}}) =0
\label{eq:con}
\end{equation}
where $\rho=\vert\psi\vert^{2} $ and the particle is assumed
to have a definite, but unknown, position with a momentum
given by
\begin{equation}
p=\nabla S. \label{eq:mom}
\end{equation}
The quantum Hamilton-Jacobi equation~(\ref{eq:hj})
contains the extra `quantum potential'
term, $Q= -\frac{\hbar^{2}} {2m} \frac{\nabla^{2}R}{R}$,
which is dependent upon the magnitude of the wavefunction
but not upon its phase. Newton's
second law is modified by the `quantum force' term and becomes
\begin{equation}
F = -\nabla{(Q+V)}.
\label{eq:for}
\end{equation}
The kinematic relation
\begin{equation}
v=\frac{j}{\rho} = \frac{p}{m},
\label{eq:vel}
\end{equation}
derived from the identification of the current density $j$ in
equation~(\ref{eq:con}), allows trajectories to be determined from
the integral curves of~(\ref{eq:mom}). Since S is determined
by~(\ref{eq:schro}), both the trajectory of an individual particle
and the evolution of its dynamical variables are determined by the
time development of ${\psi}$. Thus, in Bohm's approach, dynamical
behaviour may be expressed through, but not derived from, the
familiar apparatus of classical mechanics. The conservation
equation~(\ref{eq:con}) ensures that if the density of an ensemble
of such trajectories is, at any time, proportional to ${\rho}$,
then it was and will be so for all time. It should be emphasised
that the Bohm formulation of quantum mechanics is an exact
expression of the Schr\" {o}dinger equation and not a WKB
approximation as is sometimes suggested.

The motion of a system may be obtained either by integrating
~(\ref{eq:vel}), or in principle, by integrating the modified
Newton's equation~(\ref{eq:for}) with the initial condition
satisfying equation~(\ref{eq:mom}).  However, unlike the external
potential, the quantum potential is not a pre-assigned function of
the system co-ordinates and can only be derived from the the
wavefunction.  It is therefore essentially quantum in nature.

\section{Generalised Derivation of the Quantum Potential}
\label{sec:gender}

We generalise the above derivation of the quantum potential
by first writing the Schr\" {o}dinger equation in the form:
\begin{equation}
i\hbar\frac{\partial\psi}{\partial t} =
\left(T(\hat {p}) + V(\hat {x})\right)\psi \label{eq:schrog}
\end{equation}
where $T(\hat {p})$ and $V(\hat {x})$ are kinetic and potential
energy functions of the momentum and position operators
$\hat {p}$ and $\hat {x}$ respectively.
The substitution $\psi= Re^{iS/\hbar}$ is made recognising that
$R$ and $S$ are functions of $x$ (in the configuration space
representation) or of $p$ (in the momentum space representation).

Applying this ansatz in configuration space, with $\hat {x} = x$
and $\hat {p} = -i\hbar\nabla_{x}$,
the real and imaginary parts of the
Schr\"{o}dinger equation become:
\begin{equation}
\frac{\partial S}{\partial t} +
\Re\left(\frac{T\psi}{\psi} \right) + V(x) = 0 \label{eq:hjx}
\end{equation}
and
\begin{equation}
\frac{\partial\rho}{\partial t} -
\frac{2\rho}{\hbar}\Im\left(\frac{T\psi}{\psi}\right) = 0.
\label{eq:conx}
\end{equation}
Correspondingly, in momentum space, with $\hat {x} = i\hbar\nabla_{p}$
and $\hat {p} = p$,
the real and imaginary parts of the Schr\"{o}dinger equation are:
\begin{equation}
\frac{\partial S}{\partial t} + T(p)
+\Re\left(\frac{V\psi}{\psi} \right) = 0  \label{eq:hjp}
\end{equation}
and
\begin{equation}
\frac{\partial\rho}{\partial t} -
\frac{2\rho}{\hbar}\Im\left(\frac{V\psi}{\psi}\right) = 0.
\label{eq:conp}
\end{equation}
The symmetry between these two representations of the Schr\"
{o}dinger equation is clear. In the remaining part of this section
we show how both classical-like and quantum terms emerge from
equations~(\ref{eq:hjx}) and~(\ref{eq:hjp}) and how the quantum
terms constitute the quantum potential.  Attention is first given
to the configuration space representation.

Expanding exponentials, we write the `kinetic' term in~(\ref{eq:hjx}) as:
\begin{equation}
\Re\left(\frac{\psi^{*}T\psi}{\rho} \right) =
\Re\left(\frac{R(1-\frac{iS}{\hbar}-...)T(\hat{p})
R(1+\frac{iS}{\hbar}-...)}{\rho} \right)  \label{eq:hjxexp}
\end{equation}
If $T(\hat{p})$ is a general but regular function of $\hat{p}$, it may be
expanded as a power series in $\hat {p} = -i\hbar\nabla_{x}$.
The `kinetic' term may then be separated into the sum of two parts:
\begin{equation}
\Re\left(\frac{T\psi}{\psi} \right) =
T_{\hbar}(x) +  T_{0}(x). \label{eq:sumT}
\end{equation}
(See Appendix.) $T_{\hbar}(x)$ is an expansion in even positive
powers of $\hbar$ and so tends to zero as $\hbar$ tends to zero. On
the other hand,
\begin{equation}
T_{0}(x) =  T(\nabla_{x}S) \label{eq:Tc}
\end{equation}
is independent of $\hbar$ and identifies $p=\nabla_{x}S$.

The same line of argument
allows the `potential' term of the Hamilton-Jacobi equation~(\ref{eq:hjp})
to be separated into:
\begin{equation}
\Re\left(\frac{V\psi}{\psi} \right) =
V_{\hbar}(p) +  V_{0}(p) \label{eq:sumV}
\end{equation}
$V_{\hbar}(p)$ is in general an expansion in even positive powers
of $\hbar$. On the otherhand,
\begin{equation}
V_{0}(p) =  V(-{\nabla_{p}S}) \label{eq:Vc}
\end{equation}
is independent of $\hbar$ and identifies $x=-\nabla_{p}S$.

In the configuration representation, it is the
simple quadratic monomial form of the kinetic energy operator:
\begin{equation}
T(\hat{p}) = \hat{p}^{2}/{2m} \label{eq:Tp}
\end{equation}
that leads to the simple one term expression:
\begin{equation}
T_{\hbar}(x) =  -\frac{\hbar^{2}} {2m} \frac{\nabla_{x}^{2}R}{R},
\label{eq:Tqxx}
\end{equation}
known as the quantum potential, while
\begin{equation}
T_{0}(x) = \frac{\left({\nabla_{x}S} \right)^{2}}{2m},
\label{eq:Tcxx}
\end{equation}
has the same form as the classical kinetic energy.

In the momentum representation, the quantum potential does not, in
general, have the simple one term form of the usual configuration
space representation. (See Appendix). The free particle aside,
there are other exceptions to this rule. For the linear potential
$V(x)\sim x$, there is no quantum potential in the momentum
representation. For the harmonic oscillator potential $V(x)=\frac
{1}{2}m{\omega}^{2}$, the quantum potential has the simple form:
\begin{equation}
V_{\hbar}(p) =  -\frac {\hbar^{2}}{2}m{\omega}^{2}
\frac{\nabla_{p}^{2}R}{R},
\label{eq:Vqpp}
\end{equation}
while
\begin{equation}
V_{0}(p) = \frac{\left({\nabla_{p}S} \right)^{2}}{2}m{\omega}^{2}
\label{eq:Vcpp}
\end{equation}
has the same form as the classical potential energy. These
equations have the same form as~(\ref{eq:Tqxx})
and~(\ref{eq:Tcxx}), being likewise generated by a quadratic
monomial form.

Potentials, $V(x)$, of higher polynomial order generate a series of
energy terms contributing to the total quantum potential in the momentum
representation.  These include the sum of products of
derivatives of $R$ and $S$ with respect to $p$; all terms
having even powers of $\hbar^{2}$.  For example,
in the monomial case of $V(x)=x^{4}$, the quantum potential in
the momentum representation is:
\begin{eqnarray}
V_{\hbar}(p) && \ = \frac{{\hbar}^{4}}{R}{\frac {\partial
^{4}R}{\partial {p}^{4}}}
\cr -6\,\frac {{\hbar}^{2}}{R}
\left (\frac {\partial ^{2}R}{\partial {p}^{2}} \right )
\left (\frac {\partial S}{\partial p} \right )^{2}
&&-12\,\frac {{\hbar}^{2}}{R}
\left (\frac {\partial R}{\partial p} \right )
\left (\frac {\partial S}{\partial p} \right )
\frac {\partial ^{2}S}{\partial {p}^{2}}
\cr -3\,{\hbar}^{2}\left (\frac {\partial ^{2}S}{\partial {p}^{2}} \right )^{2} \break
&&-4\,{\hbar}^{2}\left (\frac {\partial S}{\partial p} \right )
\frac {\partial ^{3}S}{\partial {p}^{3}}
\label{eq:Vq4}
\end{eqnarray}
and the external potential is:
\begin{equation}
V_{0}(p) = \left (- \frac {\partial S}{\partial p} \right )^{4}
\label{eq:Vc4}
\end{equation}
The composition of the quantum potential in the momentum representation
is therefore, in general, much harder to evaluate and
interpret than in the configuration space representation.

\section{The symplectic structure of the causal interpretation}
\label{sec:symplc}

We build upon the previous section by expressing the
dynamic~(\ref{eq:for}) and kinematic~(\ref{eq:vel}) relations of the
causal interpretation in
Hamiltonian form using an effective Hamiltonian.

From equations~(\ref{eq:hjx}),~(\ref{eq:sumT}) and~(\ref{eq:Tc}) of the
configuration space representation, the causal trajectory moves under
the influence of the effective Hamiltonian
\begin{equation}
H_{x} = T(p) +  T_{\hbar}(x) + V(x), \label{eq:Hx}
\end{equation}
where the identification $p=\nabla_{x}S$ is understood.  The kinematic and
dynamic equations of motion for the trajectory emerge directly from
Hamilton's equations:
\begin{equation}
\dot x =
\left[\frac{\partial {H_{x}}}{\partial p}\right]_{p=\nabla_{x}S} =
\left[\frac{\partial {T(p)}}{\partial p}\right]_{p=\nabla_{x}S}
\label{eq:xdotx}
\end{equation}
and
\begin{equation}
\dot p =
-\frac{\partial {H_{x}}}{\partial x} =
-\frac{\partial {\left( T_{\hbar}(x) + V(x)\right)}}{\partial x}.
\label{eq:pdotx}
\end{equation}

Similarly, from equations~(\ref{eq:hjp}),~(\ref{eq:sumV}) and~(\ref{eq:Vc})
of the
momentum space representation, the causal trajectory moves under
the influence of the effective Hamiltonian
\begin{equation}
H_{p} = T(p) +  V_{\hbar}(p) + V(x),
\label{eq:Hp}
\end{equation}
where the identification $x=-\nabla_{p}S$ is understood. Using a
Legendre transformation, the Hamiltonian form of the kinematic
(\ref{eq:xdotx}) and dynamic~(\ref{eq:pdotx}) equations of motion
in configuration space may be converted to momentum space but
applied to the effective Hamiltonian~(\ref{eq:Hp}).  Thus,
\begin{equation}
\dot p =
-\left[\frac{\partial {H_{p}}}{\partial x}\right]_{x=-\nabla_{p}S} =
-\left[\frac{\partial {V(x)}}{\partial x}\right]_{x=-\nabla_{p}S}
\label{eq:pdotp}
\end{equation}
and
\begin{equation}
\dot x =
\frac{\partial {H_{p}}}{\partial p} =
\frac{\partial {\left(T(p) + V_{\hbar}(p)\right)}}{\partial p}.
\label{eq:xdotp}
\end{equation}

Comparison between equations~(\ref{eq:xdotx}) and~(\ref{eq:xdotp}) and
between equations~(\ref{eq:pdotp}) and~(\ref{eq:pdotx}) shows how the
precise symplectic symmetry of classical mechanics is broken by the
quantum potential terms in momentum and configuration space respectively.
Thus, in general, the trajectory $x_{x}(t)$ through
configuration space
is not the same as the trajectory $x_{p}(t)$ in momentum space.
$p_{p}(t)$ and $p_{x}(t)$ are, in general, correspondingly different.
These differences seem to reflect the complementary dispersions of
position and momentum in quantum mechanics and suggest a connection
between the breaking of symplectic symmetry by the quantum potential and
the Heisenberg uncertainty principle.

\section{Dispersion and Localisation}
\label{sec:disploc}

Insofar as the quantum potential is a uniquely quantum energy,
it is natural to enquire of the source of that energy and to try
to identify its composition. A single quantum system has
kinetic energy and potential energy.  However,
unlike single classical systems, it also has intrinsic internal
energies respectively associated with spatial localisation and
momentum dispersion.  These quantum features,
whose complementary relationship is expressed through the Heisenberg
uncertainty relation, form the mechanism by which a quantum system
maintains its non-local (ie non-point like) identity in phase space.
In the following, we show how, in the configuration space representation,
the quantum potential can expressed as the sum of localisation and
dispersion energies.  Attention
is then given to the same relationship in the momentum space representation.

\subsection{Configuration space}
\label{sec:confsp}

In order to develop the link between the quantum potential and
localisation and dispersion energies, we first write the quantum potential
in configuration space as the sum of two terms:
\begin{equation}
T_{\hbar}(x) =
\frac{-\hbar^{2}}{2mR}{\nabla_{x}}^{2}R =
{M}_{d}(x) + {\Sigma}_{l}(x) \label{eq:disp1}
\end{equation}
where
\begin{equation}
{M}_{d}(x) =
\frac{-\hbar^{2}}{8m}\left({\nabla_{x}}^{2}\ln{\rho} \right)
\label{eq:disp1a}
\end{equation}
and
\begin{equation}
{\Sigma}_{l}(x) =
\frac{-\hbar^{2}}{8m}\left(\frac{{\nabla_{x}}^{2}\rho}{\rho} \right).
\label{eq:disp1b}
\end{equation}
The meaning of the suffices is explained below.  As mentioned in
section~{\ref{sec:intro}}, Ferry and
Zhou~\cite{Ferry93} identify ${M}_{d}(x)$ as the Wigner potential whilst
themselves introducing ${\Sigma}_{l}(x)$ (which we call the Ferry-Zhou
potential) as the quantum correction
potential in their representation of the QHD equations for
semi-conductor modelling.  Expressing the quantum
potential in the form of~(\ref{eq:disp1})
also allows us to make an important
connection with the work of
Takabayasi~\cite{Tkbysi53} on the quantum mechanics of pure states on
phase space.

Takabayasi~\cite{Tkbysi53} determines
conditions on all distribution moments of the Wigner density
$f(x,p)$ in phase
space which guarantee its representing a pure state.  These conditions are
determined by projections onto configuration space and onto
momentum space.  Here we present the conditions on the second moments
of the distribution $f(x,p)$ with respect to momentum which apply for
the projection onto a one-dimensional configuration space. (The
generalisation to all space dimensions is given in~\cite{Tkbysi53}).
The probability density and the first and second
moments of the Wigner density are
given by:
\begin{equation}
\rho(x) = \int f(x,p) dp, \label{eq:mom0}
\end{equation}
\begin{equation}
P^{(1)}(x) = \int pf(x,p) dp, \label{eq:mom1}
\end{equation}
and
\begin{equation}
P^{(2)}(x) = \int p^{2}f(x,p) dp \label{eq:mom2}
\end{equation}
Defining the mean moments as $\overline {p(x)} = P^{(1)}(x)/\rho$ and
$\overline {p(x)^{2}} = P^{(2)}(x)/\rho$, the lowest order condition on the
moments is the dispersion relation:
\begin{equation}
\overline {p(x)^{2}} - {\overline {p(x)}}^2 =
\frac{-\hbar^{2}}{4}{\nabla_{x}}^{2}\ln{\rho}.  \label{eq:disp2}
\end{equation}
Thus, ${M}_{d}(x)$ in~(\ref{eq:disp1a}) is identified as the momentum
dispersion energy.  Since the Wigner density can be negative for quantum
systems,
the momentum dispersion can also
be negative.  In classical ensembles, the momentum dispersion is never
negative because the Liouville density in phase space is never negative.

${\Sigma}_{l}(x)$, in~(\ref{eq:disp1b}), is a measure of the local
curvature of the probability density $\rho$ and we call it the
localisation energy for the following reason: It contributes
positively to the quantum potential in regions of negative
curvature (eg near maxima) and contributes negatively in regions of
positive curvature (eg near minima) which correspond to high
spatial dispersion. The latter is in contrast to the positive
contribution to the quantum potential caused by a high momentum
dispersion. The quantum potential in configuration space is
therefore the balance between the energies of spatial and momentum
dispersion in quantum systems. This balance suggests a link between
the quantum potential and Heisenberg's position-momentum
uncertainty principle.

It is important to realise that the above process of projection
from phase space onto configuration space can be applied to a
classical ensemble of particles each subject to the corresponding
classical Hamiltonian.  Muga et al.~\cite{Muga93} show how this
leads to an internal potential in configuration space which
appears, in addition to the external potential, in the classical
Hamilton-Jacobi equation.  As in the quantum case, the internal
potential will not vanish unless the momentum dispersion energy
vanishes.  However, we note here that the quantum (internal)
potential is distinguished (at least for a pure state) by being a
function of the probability density in $x$.  In particular, the
momentum dispersion energy is dependent upon the density
distribution in $x$; a feature certainly absent from the
corresponding classical ensemble of non-interacting particles. This
inter-dependence of the momentum and configuration space
distributions clarifies the link between the quantum potential and
the Heisenberg uncertainty principle.

In the causal interpretation of quantum mechanics, the quantum
system is represented by an ensemble of particle trajectories (a
classical notion) whose density in configuration space is at all
times proportional $\rho(x,t)$.  Each trajectory is not only
subject to the external potential, $V(x)$, but also to the
`internal' quantum potential, $T_{\hbar}(x)$, which, being composed
of the internal energies $M_{d}(x)$ and $\Sigma_{l}(x)$, reflects
the internal structure of the extended quantum system.  Such
`internal' energies are absent from classical particles, because
they are points having no inner structure. It is only through
ensembles of classical particle trajectories that a representation
of the quantum system can be obtained.

We can propose an alternative interpretation that avoids reference
to classical particles and treats the quantum system as an extended
continuum.  In such a model, the quantum potential in ${\delta}{x}$
at $x$ applies to the fraction ${\rho}(x){\delta}{x}$ of the
quantum system rather than to the whole system represented as a
classical particle at $x$ with the appropriate probability
weighting. The internal energy per unit volume at $x$ is therefore
${\rho}(x)T_{\hbar}(x)$ and may be described as the internal
`quantum stress' or quantum potential density
${\gamma}_{\hbar}(x)$. By multiplying~(\ref{eq:disp1}) throughout
by ${\rho}(x)$ the quantum potential density,
${\gamma}_{\hbar}(x)$, may be expressed as the sum of momentum
dispersion and the spatial localisation energy densities:
\begin{equation}
{\gamma}_{\hbar}(x) = {{\rho}(x)}{{T}_{\hbar}(x)} = {\mu}_{d}(x) +
{\sigma}_{l}(x) \label{eq:dispd1}
\end{equation}
where
\begin{equation}
{\mu}_{d}(x) =
\frac{-\hbar^{2}}{8m}\left({\rho}{\nabla_{x}}^{2}\ln{\rho} \right)
\label{eq:dispd1a}
\end{equation}
and
\begin{equation}
{\sigma}_{l}(x) =
\frac{-\hbar^{2}}{8m}\left({\nabla_{x}}^{2}\rho \right). \label{eq:dispd1b}
\end{equation}
The energy densities, ${\mu}_{d}(x)$ and ${\sigma}_{l}(x)$,
themselves provide
information on the internal structure of the quantum system
and result from the projection of Wigner phase space representation of
the quantum system onto configuration space.

Takabayasi~\cite{Tkbysi53} gives particular attention to the
`quantum stress' $-{2}{\mu}_{d}(x)$ whose gradient appears as a
source of `quantum' momentum flow in the configuration space
momentum conservation equation. This source of momentum flow is in
addition to that generated by the gradient of the  `classical
stress' caused by the external potential.  The quantum potential
itself only appears as an additive correction to the external
potential in the equation for the total time derivative of the
momentum.  The latter expresses the dynamics of a point moving with
the configuration space momentum flow at a velocity of ${\overline
{p(x)}}/m$ and has precisely the form of equation~(\ref{eq:for}).
Here we see that in order to represent quantum dynamics, such a
point is subject to forces arising from momentum dispersion and
spatial localisation in addition to the external force.

It is easily shown, from~(\ref{eq:disp1a}) and~(\ref{eq:disp1b}),
that
\begin{equation}
{M}_{d}(x) {\geq} {\Sigma}_{l}(x),
\label{eq:gec}
\end{equation}
and thus
\begin{equation}
{\mu}_{d}(x) {\geq} {\sigma}_{l}(x).
\label{eq:gedc}
\end{equation}
The equality applies at maxima of the density ${{\rho}(x)}$, where
both energy (density) components contribute equally to the
quantum potential (density).  This
point is illustrated in the examples in section~{\ref{sec:expls}}.

\subsection{Momentum space}
\label{sec:momsp}

In the momentum representation, the quantum
harmonic oscillator is alone in having a quantum potential in the form:
\begin{equation}
V_{\hbar}(p) =
\frac{-\hbar^{2}}{2R}m{\omega^{2}}{\nabla_{p}}^{2}R =
{\Sigma}_{d}(p) + {M}_{l}(p)  \label{eq:disp1m}
\end{equation}
where
\begin{equation}
{\Sigma}_{d}(p) =
\frac{-\hbar^{2}}{8}m{\omega^{2}}\left({\nabla_{p}}^{2}\ln{\rho} \right)
\label{eq:disp1ma}
\end{equation}
and
\begin{equation}
{M}_{l}(x) =
\frac{-\hbar^{2}}{8}m{\omega^{2}}\left(\frac{{\nabla_{p}}^{2}\rho}{\rho} \right). \label{eq:disp1mb}
\end{equation}
We invoke the momentum space position dispersion relation derived
for a pure state by Takabayasi~\cite{Tkbysi53}, by projecting from phase
onto momentum space in a
manner similar to that described above for configuration space:
\begin{equation}
\overline {x(p)^{2}} - {\overline {x(p)}}^2 =
\frac{-\hbar^{2}}{4}{\nabla_{p}}^{2}\ln{\rho}.  \label{eq:disp2m}
\end{equation}
Thus, ${\Sigma}_{d}(p)$, in~(\ref{eq:disp1ma}), is identified
as the spatial dispersion energy which can be positive or negative.
${M}_{l}(p)$, in
~(\ref{eq:disp1mb}), is the momentum localisation energy term and is
proportional to the local curvature of the probability density in momentum
space.
Therefore, in the momentum representation of the quantum harmonic
oscillator,
the quantum potential receives positive contributions from
the spatial dispersion energy and from the momentum localisation energy.
The latter observation complements the similar one made for the
configuration space representation, in which the roles of position and
momentum are interchanged.  Thus, in the momentum representation of the
quantum harmonic oscillator also, is the balance of spatial and
momentum dispersion suggestive of the Heisenberg uncertainty principle.

As in the configuration space representation, a distinction may be
drawn between the trajectory and continuum interpretations of the
quantum system in the momentum representation of the quantum
harmonic oscillator. Similarly, in the latter interpretation, the
quantum potential density, ${\upsilon}_{\hbar}(p)$, can be
expressed as the sum of spatial dispersion and momentum
localisation energy densities:
\begin{equation}
{\upsilon}_{\hbar}(p) = {{\rho}(p)}{V_{\hbar}(p)} = {\sigma}_{d}(p)
+ {\mu}_{l}(p)  \label{eq:dispd1m}
\end{equation}
where
\begin{equation}
{\sigma}_{d}(p) =
\frac{-\hbar^{2}}{8}m{\omega^{2}}\left({\rho}{\nabla_{p}}^{2}\ln{\rho} \right)
\label{eq:dispd1ma}
\end{equation}
and
\begin{equation}
{\mu}_{l}(p) =
\frac{-\hbar^{2}}{8}m{\omega^{2}}\left({\nabla_{p}}^{2}\rho \right).
\label{eq:dispd1mb}
\end{equation}

In the case of potentials, $V(x)$, of higher polynomial
order than two, as shown in~(\ref{eq:Vq4}), the momentum
representation yields a multi-termed sum for the quantum
potential.  Thus decomposition of the quantum potential (density)
into the dispersion and localisation energy (densities), as
described above, is not possible in general.
Therefore, whilst, in correspondence to~(\ref{eq:gec})
and~(\ref{eq:gedc}), the inequalities:
\begin{equation}
{\Sigma}_{d}(p) {\geq} {M}_{l}(p)
\label{eq:gem}
\end{equation}
and
\begin{equation}
{\sigma}_{d}(p) {\geq} {\mu}_{l}(p)
\label{eq:gedm}
\end{equation}
always hold in the momentum representation, in general, the
dispersion and localisation energy (densities) do not sum to
give the quantum potential energy (density).

\section{Examples}
\label{sec:expls}

To illustrate the discussion of the previous sections, we
investigate the symplectic symmetry breaking role and the dispersion and
localisation energy components
of the quantum potential for two simple systems: a quantum particle
in a linear potential and the quantum harmonic oscillator.

\subsection{The linear potential}
\label{sec:linpot}

\subsubsection{Configuration space}
\label{sec:linpotcon}

In the configuration space representation, the stationary one dimensional
Schr\" {o}dinger equation for a particle of mass $m=1$
in the potential $V(x)=x/2$ is:
\begin{equation}
-\frac{1}{2}\frac{{\partial}^{2}{\psi}}{\partial x^{2}} +
\frac{x}{2}\psi = E\psi \label{eq:selinc}
\end{equation}
in units where $\hbar=1$. (For an analysis of this system in quantum
phase space see~\cite{Torres96} ).
The physically acceptable solution to this equation
is given (up to a multiplicative constant)
by the Airy function~\cite{Abram70}:
\begin{equation}
\psi{(x)}=Ai(x-2E). \label{eq:solinc}
\end{equation}
$x=2E$ is the turning
point of a classical particle trajectory of energy $E$ initially travelling
from $x=-\infty$ and being reflected by the linear potential at $x=2E$ to
return to $x=-\infty$. Since the potential is unbounded, $E$ may take on
a continuum of values.  Without loss of generality, we consider the case
$E=0$ in which the classical trajectory would be reflected at the origin
$x=0$.  The density of the wavefunction and the composition of the quantum
potential for this system are
shown in Fig.\ \ref{fig1}.
At the peaks of the
density, the momentum dispersion energy, ${M}_{d}(x)$, and the
spatial localisation energy, ${\Sigma}_{l}(x)$,
(given respectively by~(\ref{eq:disp1a}) and~(\ref{eq:disp1a}))
contribute equally to the quantum potential but become
infinite in magnitude at the minima of the density.
The quantum potential
in the configuration space representation is thus manifested as the sum
the energies of momentum dispersion and spatial localisation.
As shown in Fig.\ \ref{fig2}, the quantum potential density and its
component energy densities remain finite, with the momentum dispersion
energy density showing far less structure than the spatial localisation
energy.
The kinematic and
dynamic equations of motion for the configuration space trajectory are
(see equations~(\ref{eq:xdotx}) and ~(\ref{eq:pdotx})
\begin{equation}
\dot x = 0 \label{eq:linxdx}
\end{equation}
since $p=\nabla_{x}S=0$ for the real wavefunction and
\begin{equation}
\dot p = 0 \label{eq:linpdx}
\end{equation}
since for stationary systems the quantum potential and external
potential sum to the (constant) total energy E.  These equations
show that the causal trajectory is a stationary point in
configuration space reflecting the exact balance between the
(constant) quantum and external forces.

\subsubsection{Momentum space}
\label{sec:linpotmom}

In the momentum representation, the stationary one dimensional
Schr\" {o}dinger equation for a particle of mass $m=1$
in the potential $V(x)=x/2$ is:
\begin{equation}
\frac{p^{2}}{2}\phi + i\frac{1}{2}\frac{\partial{\phi}}{\partial p}=
E\phi \label{eq:selinp}
\end{equation}
in units where $\hbar=1$.
The solution to this equation (obtained either directly or by recognising
it as the Fourier transform of the Airy function $Ai(x-2E)$ in its
integral representation{~\cite{Abram70}})
is given (up to a multiplicative constant) by
\begin{equation}
\phi{(p)}=exp\left(2ip\left( p^{2}/6 - E \right) \right). \label{eq:solinp}
\end{equation}
Maintaining consistency with the configuration space representation, we take
$E=0$.  As shown in Section~{\ref{sec:gender}}, there is no quantum potential in the
momentum space representation of a particle in a linear potential.
The kinematic and
dynamic equations of motion for the momentum space trajectory are
(see equations~(\ref{eq:xdotp}) and ~(\ref{eq:pdotp}))
\begin{equation}
\dot p = -\frac{1}{2} \label{eq:linxdp}
\end{equation}
and
\begin{equation}
\dot x = p \label{eq:linpdp}
\end{equation}
and are consistent with the causal relation $x=-\nabla_{p}S=2E-p^{2}$.
The latter equation is also the classical equation of energy conservation for
a particle in a linear potential which, in the absence of
the quantum potential, has been retrieved from the momentum representation
of the causal interpretation of quantum mechanics.

\subsubsection{Discussion}
\label{sec:lindisc}

In this example, the decomposition of the quantum potential (density)
into the
momentum dispersion energy (density) and spatial localisation (density)
is apparent in
the configuration space representation but absent in the momentum space
representation.  In the latter, the constancy of the magnitude of the
wavefunction means that both the spatial dispersion and momentum
localisation energies are zero, though, strictly, they cannot be related to
the quantum potential in this example.

The example also illustrates how the quantum potential breaks the
symplectic symmetry of classical mechanics.  In the configuration
space representation, the quantum potential acts to cancel the
effects of the external potential thus making the trajectory
stationary. In the momentum representation, the absence of the
quantum potential means that the external potential alone acts upon
a causal trajectory, making it identical to the classical
trajectory. Furthermore, the absence of the quantum potential and
its components makes it impossible, in this case, to adopt an
energy density description of the quantum system in the momentum
representation.

\subsection{The quantum harmonic oscillator}
\label{sec:qho}

\subsubsection{Configuration space}
\label{sec:qhocon}

In the configuration space representation, the stationary one dimensional
Schr\" {o}dinger equation for a particle of mass $m=1$
in the potential $V(x)=x^{2}/2$ is:
\begin{equation}
-\frac{1}{2}\frac{{\partial}^{2}{\psi}}{\partial x^{2}} +
\frac{x^{2}}{2}\psi = E\psi \label{eq:seqho}
\end{equation}
in units where $\hbar=1$. (For an analysis of this system in quantum
phase space see~\cite{Torres93} ). The eigenfunction corresponding to the
$n^{th}$ eigenvalue, $E_{n} = (n + \frac{1}{2})$, of this equation is
well known as:
\begin{equation}
{\psi}_{n}(x) = H_{n}(x){exp \left(-\frac{x^{2}}{2}\right)} \label{eq:solqho}
\end{equation}
for $n=0,1,2,...$ where $H_{n}(x)$ is the $n^{th}$ Hermite
polynomial in x. Fig.\ \ref{fig3} and Fig.\ \ref{fig4} respectively
show the composition of the quantum potential, $T_{\hbar}(x)$, for
the ground state ${\psi}_{0}(x)$ and for the second excited state,
${\psi}_{2}(x)$.  In the ground state (Fig.\ \ref{fig3}), the
momentum dispersion energy, ${M}_{d}(x)$, is constant, showing that
the `quantum force' on causal trajectories arises solely from the
variation in the spatial localisation energy, ${\Sigma}_{l}(x)$. In
the case of the second excited state (Fig.\ \ref{fig4}), variations
in both the momentum dispersion and spatial localisation energies
contribute to the `quantum force'.  Fig.\ \ref{fig5} and Fig.\
\ref{fig6} show the finite components of the quantum potential
density for the same two states of the quantum harmonic oscillator.
Figs.\ \ref{fig3} to \ \ref{fig6} clearly illustrate the
inequalities~(\ref{eq:gec}) and~(\ref{eq:gedc}). As for the linear
potential example, Fig.\ \ref{fig6} shows that whilst
$\sigma_{l}(x)$ is oscillatory, $\mu_{d}(x)$ is only very weakly so
within the `classical domain' of the quantum system in which the
quantum potential (density) is non-negative.  Outside this domain,
the phenomenon of quantum tunnelling is manifest.  The profile of
$\mu_{d}(x)$, brings to mind Takabayasi's
observation~\cite{Tkbysi53} that `...the pressure in the
configuration space ensemble results from the momentum dispersion
of the underlying phase space ensemble just in the same manner as
the pressure of ideal gas results from the thermal motion of
molecules'.  However, as far as such analogies can be helpful, in
our work here we see that this `thermal' pressure is not alone in
contributing to the nett internal pressure (potential energy
density) of a quantum system.

The kinematic and
dynamic equations of motion for the configuration space trajectory are
(see equations~(\ref{eq:xdotx}) and ~(\ref{eq:pdotx})
\begin{equation}
\dot x = 0 \label{eq:qhoxdx}
\end{equation}
since $p=\nabla_{x}S=0$ for the real wavefunction and
\begin{equation}
\dot p = 0 \label{eq:qhopdx}
\end{equation}
since for stationary systems the quantum potential and external
potential sum to the (constant) total energy E.  As in the case of
the linear potential, these equations again show that the causal
trajectory is a stationary point in configuration space reflecting
the exact balance between the (constant) quantum and external
forces.

\subsubsection{Momentum space}
\label{sec:qhomom}

In the momentum representation, the Schr\" {o}dinger equation
corresponding to equation~(\ref{eq:seqho}) is:
\begin{equation}
\frac{p^{2}}{2}\phi -
\frac{1}{2}\frac{{\partial}^{2}{\phi}}{\partial p^{2}} =
E\phi \label{eq:seqhom}
\end{equation}
The $n^{th}$ eigenvalue, $E_{n} = (n + \frac{1}{2})$, has the eigenfunction:
\begin{equation}
{\phi}_{n}(p) = H_{n}(p){exp \left(-\frac{p^{2}}{2}\right)} \label{eq:solqhom}
\end{equation}
for $n=0,1,2,...$ where $H_{n}(p)$ is the $n^{th}$ Hermite polynomial in p.
The form and composition of the quantum potential (density) in the
momentum representation (see equations~(\ref{eq:dispd1})
and~(\ref{eq:dispd1m})) can be obtained directly from its form and
composition in configuration space by making the symbolic transformations
$x \rightarrow p$,
\begin{equation}
M_{d}(x) \rightarrow \Sigma_{d}(p),~
\Sigma_{l}(x) \rightarrow M_{l}(p)
\label{eq:qhosub}
\end{equation}
and
\begin{equation}
\mu_{d}(x) \rightarrow \sigma_{d}(p),~
\sigma_{l}(x) \rightarrow \mu_{l}(p).
\label{eq:qhosubd}
\end{equation}
With these substitutions, Figs.\ \ref{fig3} to\ \ref{fig6} give the
form and composition of the quantum potential (density) of the quantum
harmonic oscillator in the momentum representation.  Observations
complementary to those made for the configuration space representation can
then be made; though the parallel analogy of the spatial dispersion energy
density $\sigma_{d}(p)$ with the pressure of an ideal gas in inverse space
is of little value.

The kinematic and
dynamic equations of motion for the momentum space trajectories are
(see equations~(\ref{eq:xdotp}) and ~(\ref{eq:pdotp})
\begin{equation}
\dot x = 0 \label{eq:qhoxdp}
\end{equation}
since $p=\nabla_{x}S=0$ for the real wavefunction and
\begin{equation}
\dot p = 0 \label{eq:qhopdp}
\end{equation}
since the quantum potential and external potential sum to the
(constant) total energy E. Thus in the momentum space also is each
causal trajectory a stationary point reflecting the exact balance
between the (constant) quantum and external forces.

\subsubsection{Discussion}
\label{sec:qhodisc}

Contrary to the linear potential case, the quantum harmonic
oscillator does not exhibit the general classical symplectic
symmetry breaking features of the quantum potential, as the latter
exactly cancels the variation of the external potential in both the
configuration and momentum space representations.  As a result and
despite its symplectic symmetry, the causal trajectories of the
quantum harmonic oscillator are not classical in either
representation.  However, the existence of the quantum potential
(density) and its components in both configuration and momentum
space does allow adoption of an energy density continuum
description of the quantum system in both representations.

\section{Conclusion}
\label{sec:conc}

In this work, we have reviewed the causal interpretation of quantum
mechanics with particular emphasis on the composition and role of the
quantum potential.  Whilst the interpretation is normally presented in
configuration space, we have attempted to give equal emphasis to its
formulation in both configuration space and momentum space.  In doing
so, we have demonstrated how, in general, the quantum potential
breaks the symplectic
symmetry that exists between these two representations in
classical mechanics.  This has been achieved by applying Hamilton's
equations of dynamics to the effective Hamiltonian of the quantum
Hamilton-Jacobi equation.

Assuming only general polynomial forms for the kinetic energy
operator, $T({\hat p})$, and the potential energy operator,
$V({\hat x})$, the corresponding terms in the quantum
Hamilton-Jacobi equation are found, in general, to be expressible
as power series in ${\hbar}^{2}$ in the configuration space and
momentum space representations respectively. The non-$\hbar$
dependent parts of the series become the corresponding classical
energy forms if the causal identifications $p=\nabla_{x}S$ and
$x=-\nabla_{p}S$ are respectively adopted in the configuration
space and momentum space representations.  The ${\hbar}^{2}$
dependent parts of the power series constitute the quantum
potential in the two representations. This series consists of the
sum of terms containing products of derivatives of the magnitude
and phase of the wavefunction.  Only in the common case of $T({\hat
p}) \sim {\hat p}^{2}$ in configuration space and in the restricted
case of $V({\hat x}) \sim {\hat x}^{2}$ in momentum space does the
quantum potential have its familiar single term form proportional
to the second derivative of the magnitude of the wavefunction.

In the above latter two cases, the quantum potential has been shown
to be the sum of a localisation energy and a dispersion energy
through a connection with the Wigner phase space representation of
quantum mechanics. These intrinsic internal energies distinguish
the extended character of quantum systems from the point-like
character of individual classical systems from which they are
absent.  Expressing these energies in energy density form it has
been shown how the latter description complements the normal causal
description of probability weighted particle trajectories. Thus, as
in the example of the momentum representation of a quantum system
in a linear potential, the absence of the quantum potential may be
complemented by the presence of non-trivial (in this case,
classical) causal trajectories. Equally, as in the example of the
quantum harmonic oscillator, the quantum potential may be
accompanied by trivial (stationary) causal trajectories.

In summary, the quantum potential breaks the symplectic symmetry of
classical mechanics and in its energy density form complements the trajectory
description of the causal interpretation quantum mechanics.
The complementary
aspects of the localisation and dispersion energy components of the quantum
potential and their explicit dependence upon the probability
density in either momentum or configuration space, can be interpreted as
a manifestation of the
Heisenberg uncertainty principle and of the extended nature of
quantum systems.  Since the quantum potential
is the only $\hbar$ dependent energy in the hamiltonian, it is
surely the essentially quantum aspect of the causal interpretation.
We have seen in this paper how, far from being a redundant concept in the
causal interpretation, analysis of the quantum potential provides insight
into the differences between the dynamics of classical and quantum mechanics.

\section*{Acknowledgement}
\label{sec:ackn}

The author wishes to acknowledge the encouragement, critical guidance and
support kindly given
by Professor B J Hiley of Birkbeck College, University of London,
during the preparation of this paper.



\appendix
\section*{}

In this appendix we show how the full series expansions of
equations (\ref{eq:sumT}) and~(\ref{eq:sumV}) naturally separate
into $\hbar$-dependent and $\hbar$-independent parts. The approach
may also be applied to the imaginary part functions in
equations~(\ref{eq:conx}) and~(\ref{eq:conp}). The analysis is
explicitly given for the kinetic energy operator, $T(\hat{p})$, but
follows through in an obviously similar manner for the potential
energy operator, $V(\hat{x})$. Thus, in the following, references
to $T(\hat{p})$ should be applied to $V(\hat{x})$ with the
appropriate transformation of the independent variable.

It is assumed that $T(\hat{p})$
is any real linear function of monomials in the orthogonal components
of $\hat{p}$ which span $L$-dimensional momentum space:
\begin{equation}
T(\hat{p}) =
\sum_{l}^{L}\sum_{m}^{M_{l}}a_{lm}{\hat{p}}_{l}^m.
\label{eq:amon}
\end{equation}
Thus, there are no terms containing products of different components of
the momentum and $T(\hat{p})$ is a sum of energies, uniquely
associated with each momentum component, which
can be separately analysed. From here on we therefore limit
the analysis to one
momentum component and drop the dimensional index $l$.

The left-hand side of equation~(\ref{eq:sumT}) may be expressed as:
\begin{equation}
\Re\left(\frac{T\psi}{\psi} \right) =
\frac{\Re\left(e^{-iS/\hbar}T(\hat{p})
Re^{iS/\hbar} \right)}{R}
\label{eq:are}
\end{equation}
where $\psi= Re^{iS/\hbar}$.  Making the identification
$\hat{p}=-i\hbar\frac{\partial}{\partial{x}}$ and expanding
each of the exponentials
as a power series gives
\begin{eqnarray}
\Re\left(\frac{T\psi}{\psi}\right) =
&&{\frac{1}{R}}
\sum_{j,k=0}^{\infty} \sum_{m}^{M} a_{m} (-1)^{j+m}
\Re(i^{j+k+m}) \nonumber\\
&&\times {\hbar}^{m-k-j} \frac{S^j}{(j!k!)}
\frac{{\partial}^{m}({R}{S}^k)}{{\partial}{x}^{m}}
\label{eq:arex}
\end{eqnarray}
The extraction of the real part of the multiple summation means that
$j+k+m \in Z_{e}$ where $Z_{e}$ is the set of even integers.
The expression must not change when
we add an arbitrary constant to $S$, ie it must be independent of a
constant shift in phase of the wavefunction.  Thus all powers of $S$
must sum to zero.  This imposes the constraint that
only $j=0$ terms can contribute to the summation.
Equally, those parts of the partial
derivative in equation~(\ref{eq:arex}) which have non-zero powers of
$S$ as factors cannot contribute to the summation.  This means that
$m$ is a maximum upper bound on $k$, ie $k \leq m$.  However, even with
this bound the expanded partial derivative will,
in general, have parts with
$S$ as a factor.  These latter may be eliminated in practice by
temporarily setting $S=0$ thus retaining only those parts of
${{\partial}^{m}({R}{S}^k)}/{{\partial}{x}^{m}}$ which have, as
factors, first and higher order derivatives of $S$.  In summary,
\begin{eqnarray}
\Re\left(\frac{T\psi}{\psi}\right) =
&&{\frac{1}{R}}
\sum_{m}^{M} \sum_{k=0,k+m \in Z_{e}}^{m} a_{m} (-1)^{m}
(-1)^{\frac{k+m}{2}} \nonumber\\
&&\times {\hbar}^{m-k} \frac{1}{(k!)}
\left[\frac{{\partial}^{m}({R}{S}^k)}{{\partial}{x}^{m}}\right]_{S:0},
\label{eq:arexr}
\end{eqnarray}
in which the subscript ${S:0}$ denotes the process of eliminating
terms with $S$ as a factor.  The combinations of k and m values
that contribute to the above sum are shown graphically in Fig.\
\ref{fig1A}. Equation~(\ref{eq:arexr}) contains only even
non-negative positive powers of $\hbar$.  The $\hbar$-dependent
component of the summation consists of terms with positive powers
of $\hbar$ as factors:
\begin{eqnarray}
T_{\hbar}(x) = &&{\frac{1}{R}}
\sum_{m}^{M} \sum_{k=0,k+m \in Z_{e}}^{< m} a_{m} (-1)^{m}
(-1)^{\frac{k+m}{2}} \nonumber\\
&&\times {\hbar}^{m-k} \frac{1}{(k!)}
\left[\frac{{\partial}^{m}({R}{S}^k)}{{\partial}{x}^{m}}\right]_{S:0},
\label{eq:arexq}
\end{eqnarray}
and the $\hbar$-independent component consists only of terms for
which $k=m$:
\begin{equation}
T_{0}(x) = \sum_{m}^{M} a_{m}
{\left(\frac{{\partial}{S}}{{\partial}{x}}\right)}^{m}.
\label{eq:arexc}
\end{equation}
Making the identification $p=\nabla_{x}S$ in~(\ref{eq:arexc}) we
recover the one dimensional classical form of the kinetic energy
function assumed in equation~(\ref{eq:amon}) above. Thus, is the
form of equation~(\ref{eq:sumT}) established.

The demonstration of the form and content of equation~(\ref{eq:sumV})
in the momentum representation proceeds exactly as above but using
$\hat{x}=i\hbar\frac{\partial}{\partial{p}}$ and $x=-\nabla_{p}S$ in place
of the corresponding configuration representation equations.  The
resulting momentum space equations for the Hamilton-Jacobi potential
term exactly mirror~(\ref{eq:arexr}),~(\ref{eq:arexq}) and~(\ref{eq:arexc}),
except for an extra ${(-1)^{-m}}$ multiplicative term in the
$m^{th}$ term of each series.

Considering the normal form $T(\hat{p})={{\hat{p}}^{2}}/2m$,
equations~(\ref{eq:arexq}) and~(\ref{eq:arexc}), and inspection of
Fig.\ \ref{fig1A} show how the kinetic term in the configuration
space representation of the Hamilton-Jacobi equation consists of
only one $\hbar$-independent term and one $\hbar$-dependent
(quantum potential) term. This is similarly the case for the
potential term of the quantum harmonic oscillator in the momentum
representation. In general, the quantum potential term, in either
representation (but more commonly in the momentum representation)
may consist of many different terms with a series of different even
non-zero powers of $\hbar$.


\vfill \eject
\eject

\begin{figure}\protect\caption{ Energy versus position
(configuration space) plot for a quantum system of total energy
$E=0$ in the potential $V(x)=x/2$.  (Units: ${\hbar}=m=1$). The
solid line (in arbitrary units) shows the variation of the density
$\rho$.  The straight dashed line shows 0.5 x quantum potential:
${T_{\hbar}(x)}/2$. The latter is the common tangent line for the
curves of spatial localisation energy, ${\Sigma}(x)$, and momentum
dispersion energy, ${M}(x)$, near their respective maxima and
minima. }  \protect\label{fig1} \end{figure}

\begin{figure}\protect\caption{ Energy density versus position
(configuration space) plot for a quantum system of total energy
$E=0$ in the potential $V(x)=x/2$.  (Units: ${\hbar}=m=1$). The
solid line (in arbitrary units) shows the variation of the density
$\rho$.  The dashed lines respectively show show 0.5 x quantum
potential density, ${{\gamma}_{\hbar}(x)}/2$, the spatial
localisation energy density, ${\sigma}_{l}(x)$, and momentum
dispersion energy density, ${\mu}_{d}(x)$. }  \protect\label{fig2}
\end{figure}

\begin{figure}\protect\caption{ Energy versus position
(configuration space) plot for the ground state ($n=0$) of the
quantum harmonic oscillator of total energy $E=0$ in the potential
$V(x)=x^2/2$.  (Units: ${\hbar}=m=1$). The solid line (in arbitrary
units) shows the variation of the density $\rho$.  The dashed lines
respectively show show 0.5 x quantum potential,
${{T}_{\hbar}(x)}/2$, the spatial localisation energy,
${\Sigma}_{l}(x)$, and momentum dispersion energy, ${M}_{d}(x)$. }
\protect\label{fig3} \end{figure}

\begin{figure}\protect\caption{ Energy density versus position
(configuration space) plot for the ground state ($n=0$) of the
quantum harmonic oscillator of total energy $E=0$ in the potential
$V(x)=x^2/2$. (Units: ${\hbar}=m=1$). The solid line (in arbitary
units) shows the variation of the density $\rho$.  The dashed lines
respectively show show 0.5 x quantum potential density,
${{\gamma}_{\hbar}(x)}/2$, the spatial localisation energy density,
${\sigma}_{l}(x)$, and momentum dispersion energy density,
${\mu}_{d}(x)$. }  \protect\label{fig4} \end{figure}

\begin{figure}\protect\caption{ Energy versus position
(configuration space) plot for the second excited state ($n=2$) of
the quantum harmonic oscillator of total energy $E=5/2$ in the
potential $V(x)=x^2/2$.  (Units: ${\hbar}=m=1$). The solid line (in
arbitary units) shows the variation of the density $\rho$.  The
dashed lines respectively show show 0.5 x quantum potential,
${{T}_{\hbar}(x)}/2$, the spatial localisation energy,
${\Sigma}_{l}(x)$, and momentum dispersion energy, ${M}_{d}(x)$. }
\protect\label{fig5} \end{figure}

\begin{figure}\protect\caption{ Energy density versus position
(configuration space) plot for the second excited state ($n=2$) of
the quantum harmonic oscillator of total energy $E=5/2$ in the
potential $V(x)=x^2/2$.  (Units: ${\hbar}=m=1$). The solid line (in
arbitary units) shows the variation of the density $\rho$.  The
dashed lines respectively show show 0.5 x quantum potential
density, ${{\gamma}_{\hbar}(x)}/2$, the spatial localisation energy
density, ${\sigma}_{l}(x)$, and momentum dispersion energy density,
${\mu}_{d}(x)$. }  \protect\label{fig6} \end{figure}

\begin{figure}\protect\caption{ Plot showing the
combinations of k and m values that contribute to the kinetic and
potential energy terms respectively in the configuration and
momentum space representations of the quantum Hamilton-Jacobi
equation. See equation~(\protect\ref{eq:arexr}) in the Appendix.
The circles show combinations that can contribute to the
Hamilton-Jacobi kinetic and potential energy terms. Lines of
constant even powers of $\hbar$ are shown dotted. $k=0$ is the line
of non-$S$ dependent combinations and $k=m$ is the line of terms
contributing to the $\hbar$-independent kinetic or potential
energy.}
\protect\label{fig1A} \end{figure}


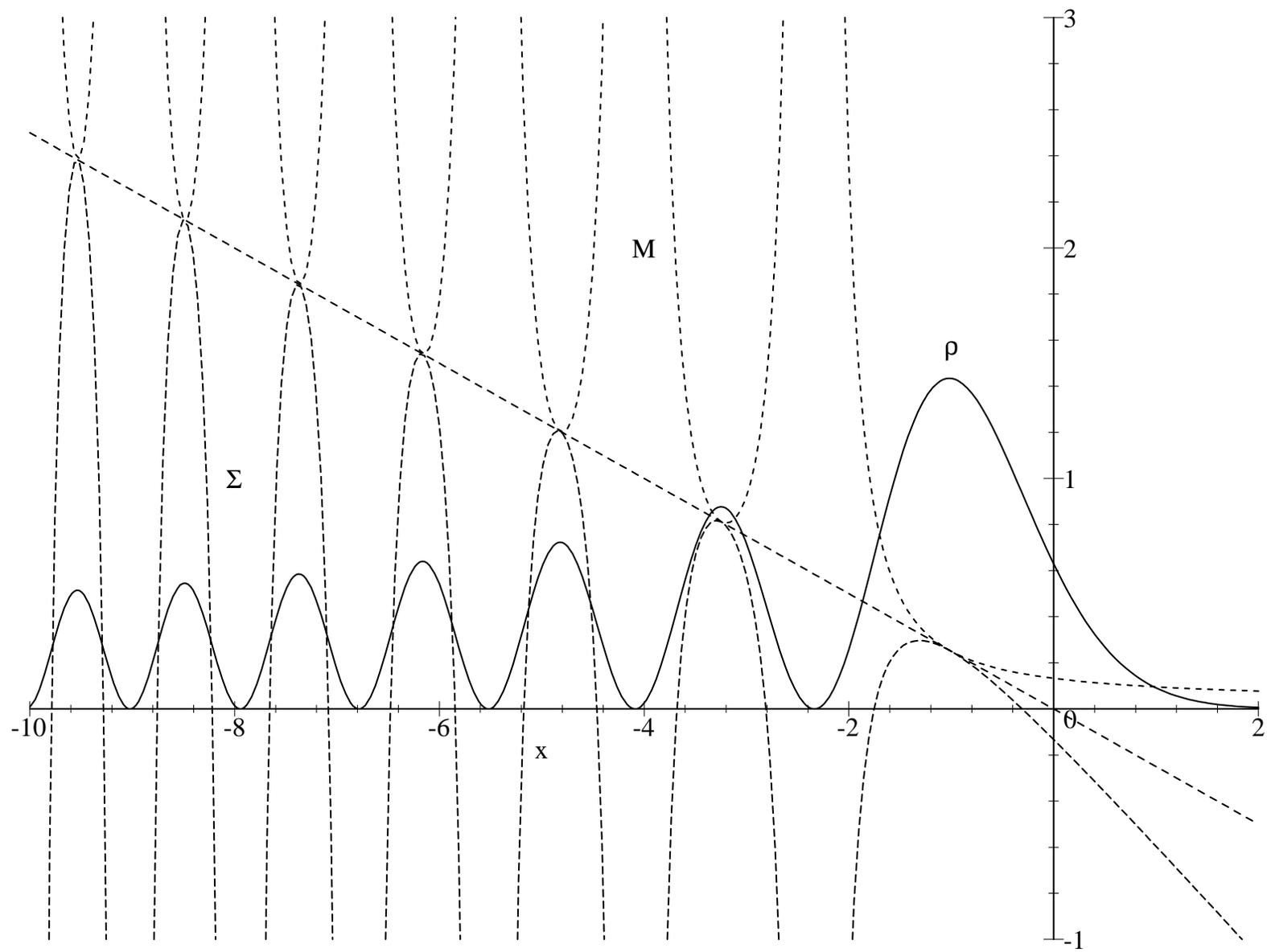
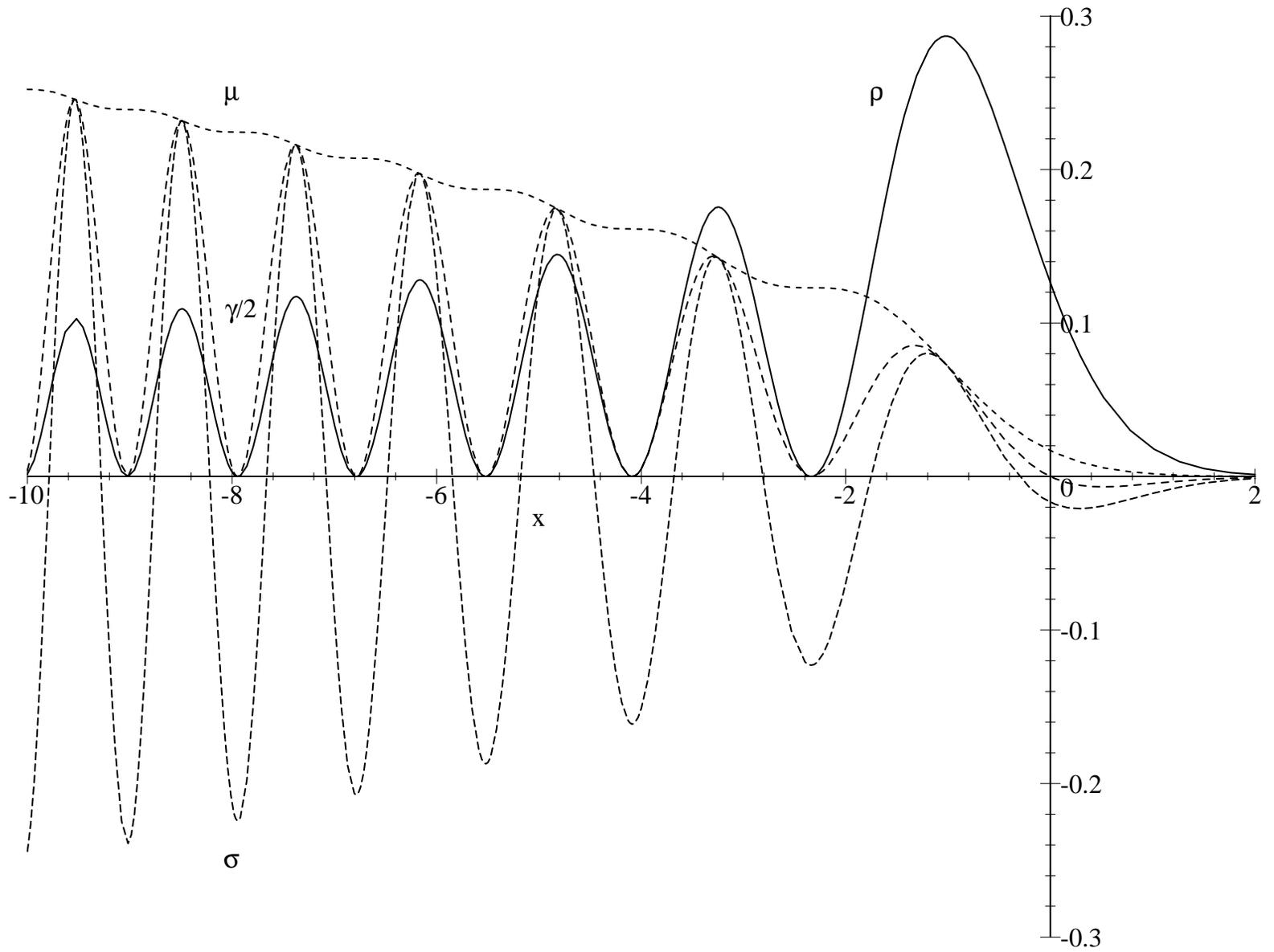
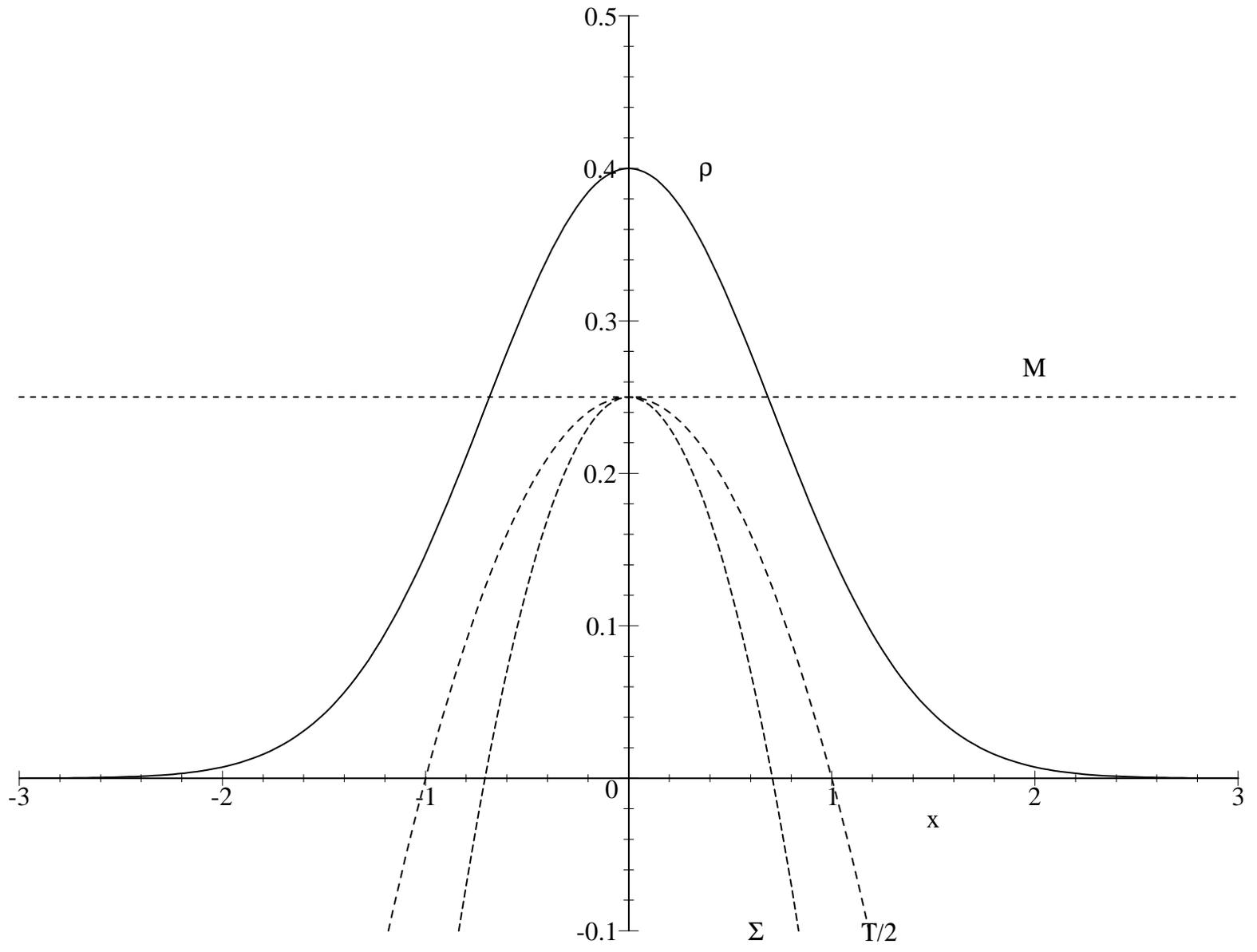
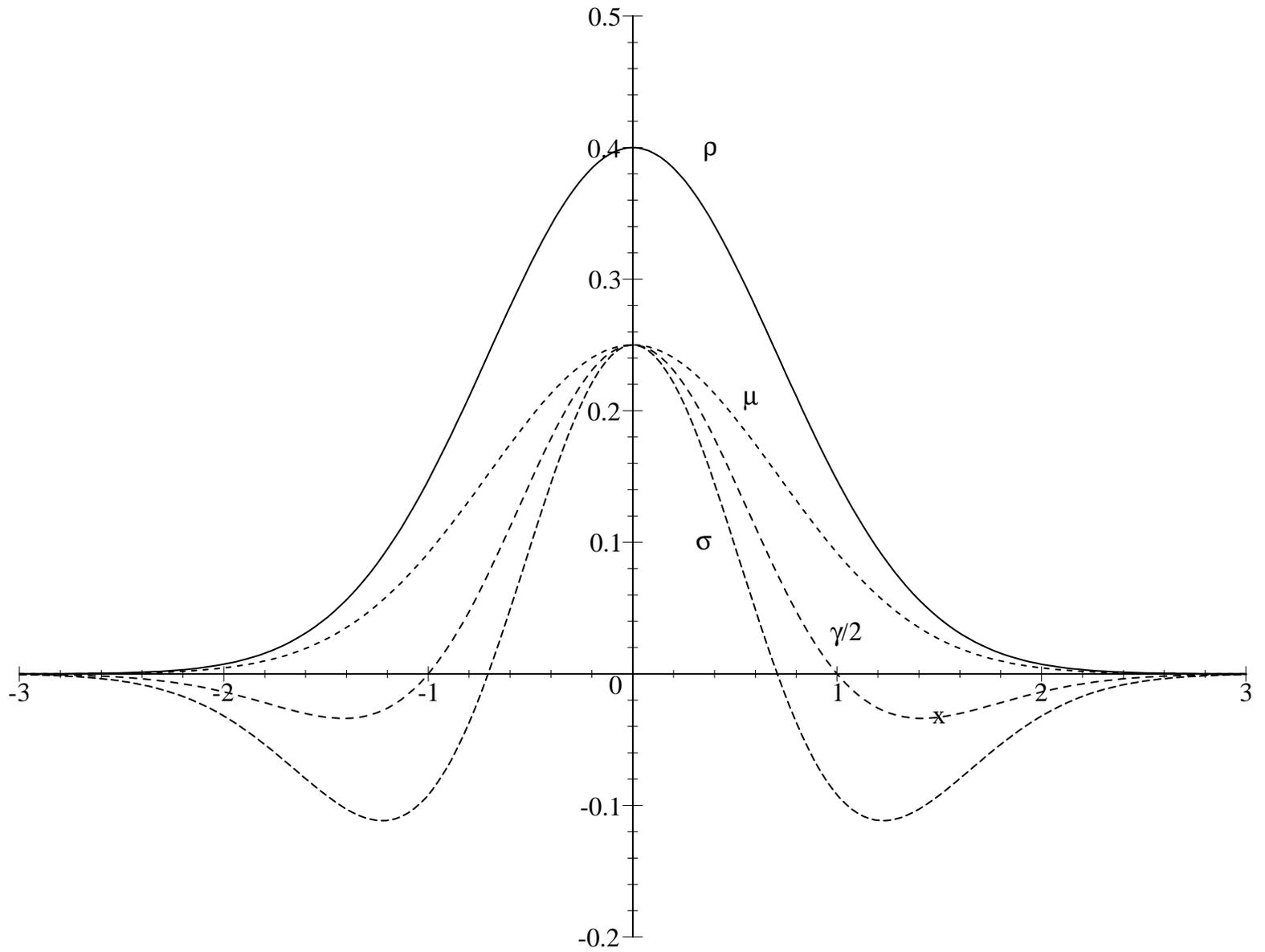
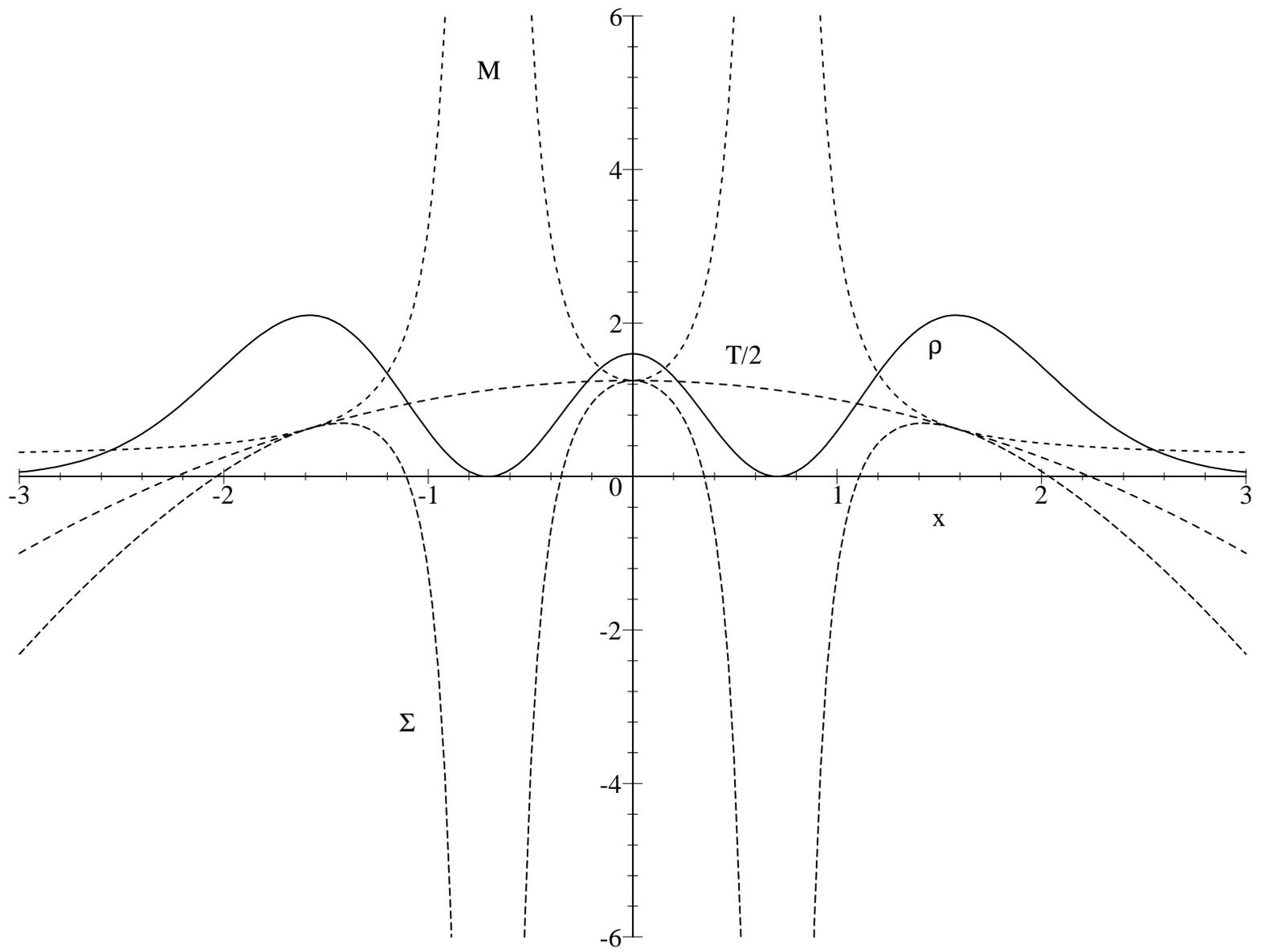
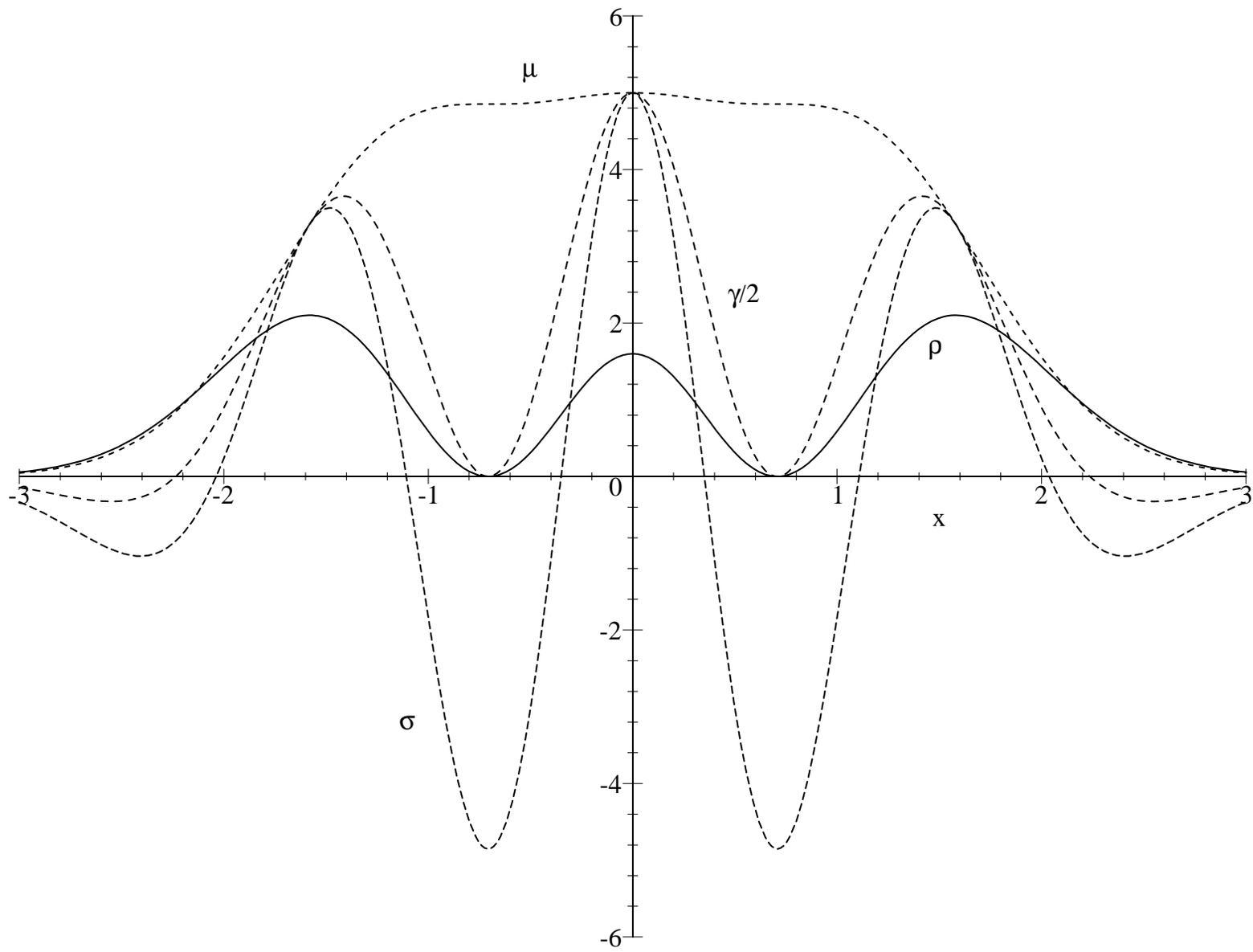
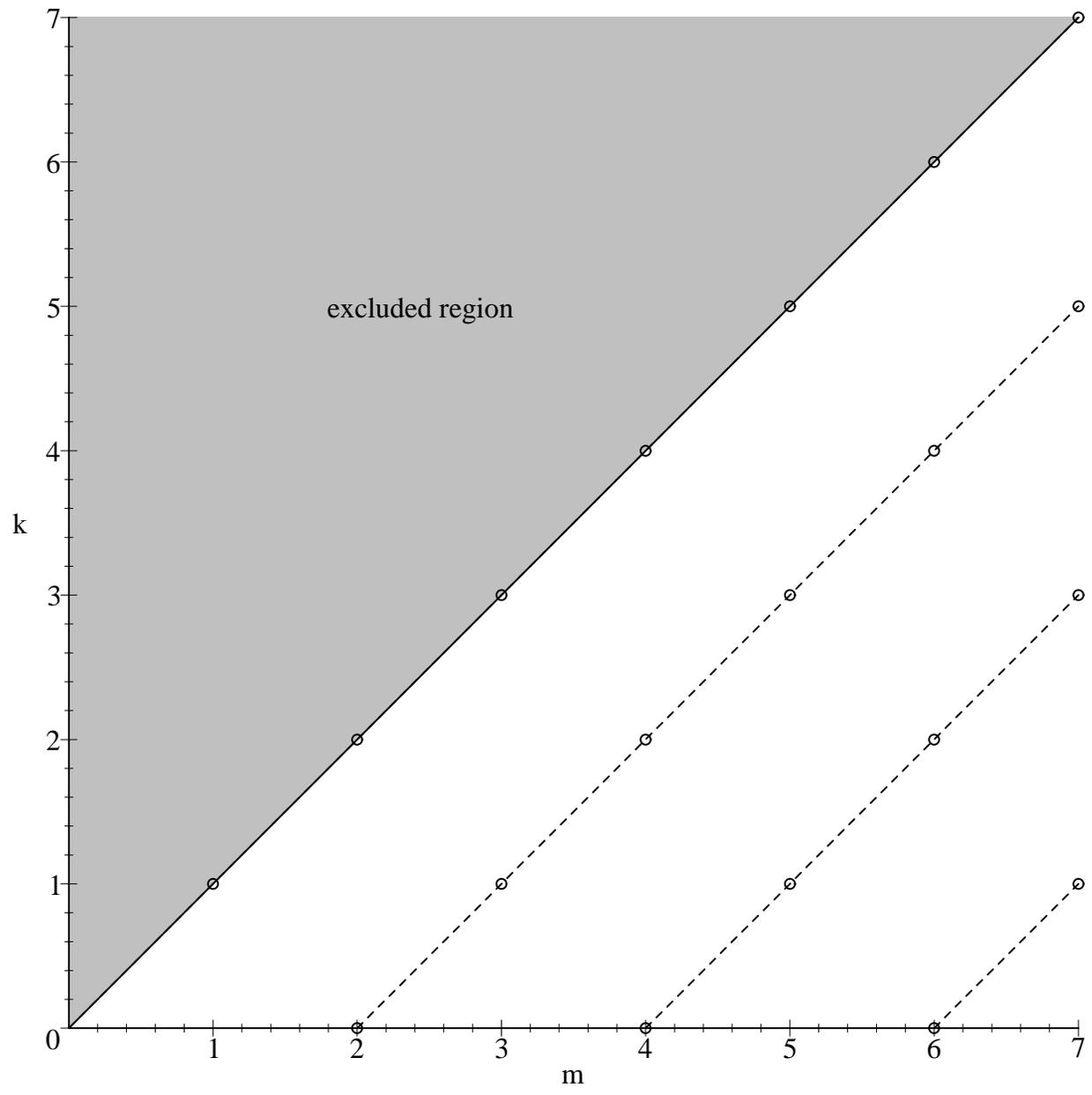
\end{document}